\documentclass[prl,aps,twocolumn,showpacs,amsmath,amssymb]{revtex4}

\usepackage{graphicx}
\usepackage{color}

\newcommand{\be}{\begin{equation}}
\newcommand{\ee}{\end{equation}}

\newcommand{\Hb}{\protect{{\bf H}}}
\newcommand{\Eb}{\protect{{\bf E}}}
\newcommand{\Pb}{\protect{{\bf P}}}
\newcommand{\Mb}{\protect{{\bf M}}}
\newcommand{\Bb}{\protect{{\bf B}}}
\renewcommand{\Im}{\text{Im}}
\renewcommand{\Re}{\text{Re}}

\newcommand{\ket}[1]{\ensuremath{\left|#1\right>}}

\begin{document}

%*************************************************************************
\title{Tunable negative refraction without absorption

via 
electromagnetically induced chirality}
%************************************************************************
\author{J\"urgen K\"astel and Michael Fleischhauer}
\affiliation{Fachbereich Physik, Technische Universit\"{a}t
Kaiserslautern, D-67663 Kaiserslautern,
Germany}

\author{Susanne F. Yelin}
\affiliation{Department Of Physics, University of Connecticut,
Storrs, Connecticut 06269, USA}
\affiliation{ITAMP, Harvard-Smithsonian Center for Astrophysics,
Cambridge, Massachusetts 02138, USA}

\author{Ronald L. Walsworth}
\affiliation{Harvard-Smithsonian Center for Astrophysics and Department
of Physics, Harvard University, Cambridge, Massachusetts 02138, USA}

\date{\today}

\begin{abstract}
We show that negative refraction with minimal absorption can be 
obtained by means of quantum interference effects similar to 
electromagnetically induced transparency. Coupling a magnetic dipole 
transition coherently with an
electric dipole transition leads to electromagnetically induced 
chirality, which can provide negative refraction without requiring 
negative permeability, and also suppresses absorption.
This technique allows negative refraction in the optical regime
at densities where the magnetic susceptibility is still small and
with refraction/absorption ratios that are orders of magnitude larger
than those achievable previously. Furthermore, the value of the refractive
index can be fine-tuned via external laser fields, which is essential
for practical realization of sub-diffraction-limit imaging.
\end{abstract}

%***********************************************************************
\maketitle
%%%%%%%%%%%%%%%%%%%%%%%%%%%%%%%%%%%%%%%%%%%%%%%%%%%%%%%%%%%%%
%%%%%%%%%%%%%%%%%%%%%%%%%%%%%%%%%%%%%%%%%%%%%%%%%%%%%%%%%%%%%%%%%%%%%%%%%%
Negative refraction of electromagnetic radiation
\cite{Veselago68} is currently a very active area of
research, driven by goals such as the development of a
``perfect lens'' in which imaging resolution is not limited by
electromagnetic wavelength \cite{Pendry00}. Despite remarkable recent
progress in demonstrating negative refraction using technologies such
as meta-materials \cite{Pendry99,Smith00,Yen04,Linden04} and photonic
crystals \cite{Parimi04,Berrier04,Lu05}, a key challenge remains the
realization of negative refraction
\textit{without absorption}
which is particularly important in the optical regime.
Here we propose a promising new
approach to this problem: the use of quantum interference effects
similar to electromagnetically induced transparency (EIT)
\cite{Fleischhauer05} to suppress absorption and induce chirality
\cite{Scully-chirality} in an ensemble of radiators (atoms,
molecules, quantum dots, excitons, etc.).

Early proposals for negative refraction required media with both
negative permittivity and permeability ($\varepsilon,\mu<0$) in the
frequency range of interest.  In the optical regime, however, it is
difficult to realize negative permeability with low loss, since
typical transition magnetic dipole moments ($\mu_a$) are smaller than
transition electric dipole moments ($d_a$) by a factor of the order
of the fine structure constant $\alpha\sim \frac{1}{137}$. As a consequence
the magnitude of magnetic susceptibilities $\chi_m$ are much smaller
than that of electric susceptibilities $\chi_e$: $|\chi_m| 
\sim\left(\mu_a/d_a\right)^2 |\chi_e|
\sim \left(1/137\right)^2 |\chi_e|$,
where $\varepsilon=1+\chi_e$ and $\mu=1+\chi_m$. Recently, Pendry
suggested an elegant way to alleviate this problem
\cite{Pendry04} by using a chiral medium, i.e., a medium in which
the electric polarization
$\Pb$ is coupled to the free-space magnetic field component $\Hb$ of
an incident optical-frequency electromagnetic wave
and the magnetization $\Mb$ is coupled to the electric field component $\Eb$:
\be
\begin{split}
\Pb=& \varepsilon_0\chi_{e}\Eb+\frac{\xi_{EH}}{c}\Hb, \\
\Mb=& \frac{\xi_{HE}}{c\mu_0}\Eb+\chi_{m}\Hb.
\end{split}
\label{Pendrychiral}
\ee
Here $\xi_{EH}$ and $\xi_{HE}$ are the chirality coefficients, which
in general can be complex.
These terms lead to additional contributions to the refractive index
% [which in a non-chiral medium is just $n=\sqrt{\varepsilon\mu}$]:
%
%
\be
n = \sqrt{\varepsilon\mu
-\frac{(\xi_{EH}+\xi_{HE})^2}{4}}
+\frac{i}{2}(\xi_{EH}-\xi_{HE}).
\label{refind}
\ee
As Pendry noted, such a chiral medium allows $n<0$  without requiring
negative permeability if there is a
positive imaginary part of $(\xi_{EH}-\xi_{HE})$ of sufficiently
large magnitude.  For example, choosing the phases of the complex
chirality coefficients such that $\xi_{EH}=-\xi_{HE}=i\xi$, with
$\xi,\varepsilon,\mu>0$, the index of refraction becomes
\be
n=\sqrt{\varepsilon\mu}-\xi,
\ee
and $n<0$ when $\xi>\sqrt{\varepsilon\mu}$. Because chirality
coefficients typically scale as
$\xi_{EH},\xi_{HE}\sim\frac{\mu_a}{d_a}\chi_e\sim\alpha\chi_e$
relative to the electric susceptibility, there is only one factor of
$\alpha$ suppression as compared to $\chi_m\sim\alpha^2\chi_e$.
Nevertheless, the use of a chiral medium to achieve
negative refraction in the optical regime still faces the demanding
requirement of minimizing loss while realizing $|\xi_{EH}|, |\xi_{HE}|\sim1$.
In the following we describe how quantum interference effects similar
to EIT allow this requirement to be met and also enable fine tuning of the
refractive index by means of external laser fields.

\begin{figure}[tb]
      \begin{center}
        \includegraphics[width=8cm]{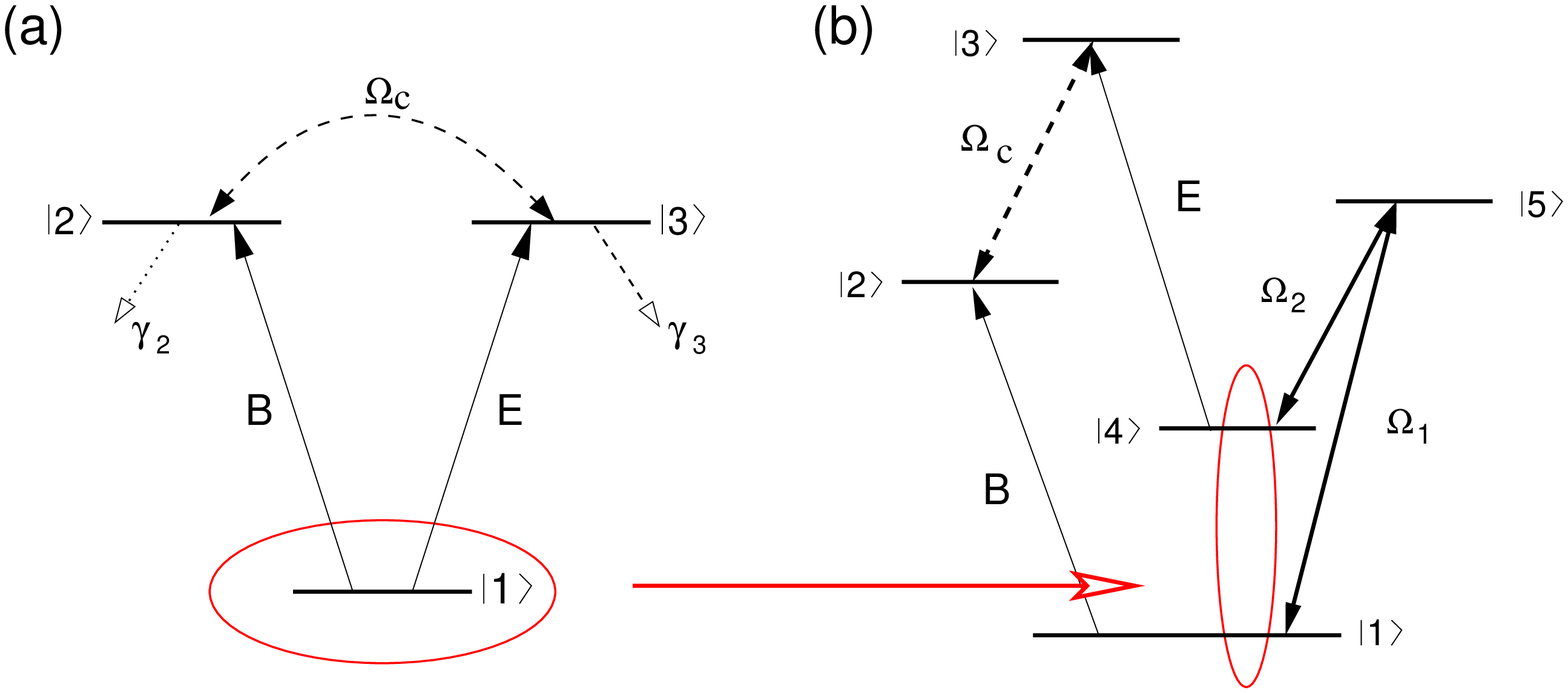}
        \caption{(color online) (a) Simplistic three-level system for 
electromagnetically-induced chiral
negative refraction.
(b) Modified level scheme for realistic media. The dark state created
by the two-photon resonant Raman coupling of levels $|1\rangle$ and 
$|4\rangle$ takes over the role of level $|1\rangle$ in 
Fig.\ref{fig1}(a). In both level schemes, $\Eb, \Bb$ are the 
electric, magnetic components of the probe field.
}
        \label{fig1}
      \end{center}
\end{figure}

To introduce concepts underlying electromagnetically induced chiral
negative refraction, we begin with the simplistic three-level system
shown in Fig.~\ref{fig1}(a), which we will later modify to a more
realistic scheme (see Fig.~\ref{fig1}(b)). As seen in Fig.~\ref{fig1}(a),
state $|1\rangle$ is coupled by an electric dipole (E1) transition to
state $|3\rangle$ and by a magnetic dipole (M1) transition to state
$|2\rangle$. The resonance frequency
of the $|1\rangle-|2\rangle$ transition is nominally identical to
that of the $|1\rangle
-|3\rangle$ transition, such that the electric ($\Eb$) and magnetic
($\Bb$) components
of the probe field %(the electromagnetic wave that experiences
% negative refraction)
can couple efficiently to the corresponding transitions.
Furthermore, there is a strong resonant coherent field coupling
the E1 transition
between states $|2\rangle$ and $|3\rangle$
with Rabi-frequency $\Omega_c$. Thus, the parities in this
system would be $\ket{1}$ - even, $\ket{2}$ - even, $\ket{3}$ - odd 
or vice versa.
An analogous parity
distribution will apply to the more realistic system of Fig.~\ref{fig1}(b).
  Since the
$|2\rangle-|1\rangle$ transition is magnetic, level $|2\rangle$
can be considered meta-stable with a magnetic-dipole decay rate
$\gamma_2 \sim (\mu_a/d_a)^2\gamma_3 \sim (1/137)^2 \gamma_3$.

The scheme of Fig.~\ref{fig1}(a) has similarities to schemes in
resonant nonlinear optics based on EIT
\cite{Harris-NLO-EIT,Stoicheff-Hakuta}. For such
schemes it is well known that there is destructive interference for the
absorption and constructive interference
for the dispersive cross-coupling between the $|1\rangle-|3\rangle$ and
$|1\rangle-|2\rangle$ transitions in the limit $\gamma_2\ll\gamma_3$
and if the transition $|1\rangle - |2\rangle$ is two-photon
resonant with the probe field $E$ and the coupling field $\Omega_c$
\cite{Fleischhauer05}.
A similar set-up has been discussed recently by Oktel and M\"ustecaplioglu
\cite{Oktel04} for the purpose of generating negative refraction
by means of a large negative permeability $\mu$. However, they did
not examine induced chirality and its effect on the index of
refraction, which as we show below, is the most important contribution.

To enable electromagnetically-induced chiral negative refraction in
realistic media, we modify the simplistic scheme of Fig.~\ref{fig1}(a)
to satisfy three criteria: (i) $\Omega_c$ must be an $ac$ field, so
that its phase can be adjusted to induce chirality; (ii) there must
be high-contrast EIT for the probe field; and (iii) the energy level
structure must be appropriate for media of interest (atoms,
molecules, excitons, etc.). In Fig.~\ref{fig1}(b) we show one example of
a modified energy level structure that meets the above criteria. This
modified scheme employs  strong coherent Raman coupling by two
coherent fields with complex Rabi-frequencies
$\Omega_1$ and $\Omega_2$ and carrier frequencies $\omega_1$ and
$\omega_2$, which creates a dark superposition of
states $|1\rangle$ and $|4\rangle$.  This dark state takes over the role of the
ground state $|1\rangle$ in the three-level scheme of
Fig.~\ref{fig1}(a), such that the electric component of the probe field
has a transition from the dark state to state $|3\rangle$ and the
magnetic component
a transition to level $|2\rangle$.  Therefore, the modified scheme
remains effectively three-level (consisting of $|2\rangle$,
$|3\rangle$, and the dark state), but with sufficient additional
degrees of freedom under experimental control to allow the above
three criteria to be satisfied.

The coherent preparation of dark
states by external laser
fields is a well established technique realized in many systems.
Existence of a dark state requires only two-photon resonance and is
largely insensitive to the intensities of the laser fields.
In the scheme of Fig.~\ref{fig1}(b), coupling of the probe field to a
dark state greatly enhances the freedom of choice of energy levels and
operating conditions, and thus makes the scheme applicable to
realistic systems.
%%%%%%%%%%%%%%%%
% For example, the
% transition frequencies  and selection
% rules can be chosen in such a way that the $\Eb$ and $\Bb$ components of
% the probe field couple predominantly to the transitions
% $|4\rangle-|3\rangle$ and $|1\rangle-|2\rangle$, respectively.
%%%%%%%%%%%%%%%%
The coupling between
the two excited states $|2\rangle$ and $|3\rangle$ is now an $ac$ coupling
with carrier frequency $\omega_c$ and Rabi frequency $\Omega_c$ which can
be chosen complex.  By varying the phase of $\Omega_c$ the phase of
$\xi_{EH}$ and $\xi_{HE}$ can be controlled.
To ensure degeneracy between the electric and magnetic carrier frequencies,
$\omega_1-\omega_2=\omega_c$ must hold.  We note that the possibility
of cross-coupling
the electric and magnetic field components of a probe field without a
common level coupled to both components was recently suggested by
Thommen and Mandel  \cite{Thommen-PRL-2006} for meta-stable neon with a
far-infrared probe field.
However, their scheme exploits neither
dark states nor chirality and thus the achieved ratio of
negative refraction to absorption is small \cite{footnote}.
Finally, since the EIT transition in the level scheme of
Fig.~\ref{fig1}(b) is predominantly between levels $|4\rangle$
and $|2\rangle$, which can have comparable energies, the two-photon
resonance can remain narrow even in the presence of
one-photon-transition broadening
mechanisms; and thus high-contrast EIT (i.e., minimal absorption of
the probe field) is
possible.

%%%%%%%%%%%%%%%%%%%%%%%%%%%%%%%%%%%%%%%%%%%%%%%%%%%%%%%%%%%%%%%%%%%%%%%%%%
%%%%%%%%%%%%%%%%%%%%%%%%%%%%%%%%%%%%%%%%%%%%%%%%%%%%%%%%%%%%%%%%%%%%%%%%%%

\begin{figure}[tb]
      \begin{center}
        \includegraphics[width=8cm]{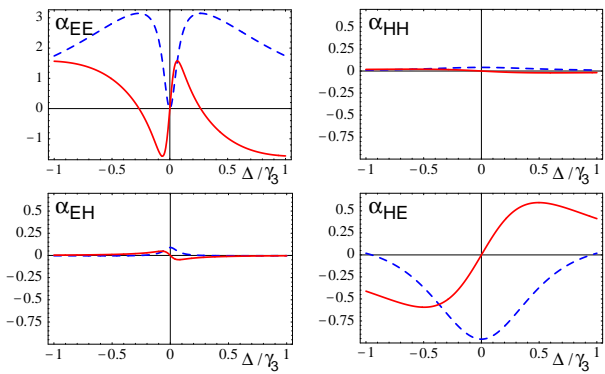}
        \caption{(color online) Real ({\color{red} solid}) and 
imaginary ({\color{blue}
dashed}) parts of the electric ($\alpha_{EE}$) and
magnetic ($\alpha_{HH}$) polarizabilities as well as
the chirality parameters ($\alpha_{HE}$,
$\alpha_{EH}$) in arbitrary but the same units.
% , as a function of
% probe field detuning $\Delta=-\Delta_E=-\Delta_B$ relative to the
% radiative decay rate $\gamma_3$ from level $|3\rangle$. Here,
% $\Delta_E=\omega_{34}-\omega_{\rm probe}$ and
% $\Delta_B=\omega_{21}-\omega_{\rm probe}$, where $\omega_{\rm probe}$
% is the probe-field frequency and $\omega_{\mu\nu}$ is the transition
% frequency between levels $|\mu\rangle$ and $|\nu\rangle$. All
% coupling fields are resonant.
}
        \label{fig2a}
      \end{center}
\end{figure}

For an ensemble of radiators described by the level scheme
of Fig.~\ref{fig1}(b), we calculate the real and imaginary
parts of the refractive index in the linear-response-limit.
State $|2\rangle$ is assumed to be
metastable since it couples to $|1\rangle$ only via a magnetic dipole
transition, which as we show below allows absorption losses to be
minimized. Similarly, state $|4\rangle$ is assumed to be stable.
To model realistic situations we take into account broadening $\gamma_P$ of the
intrinsically very narrow
magnetic transition $|1\rangle -|2\rangle$ which could be due to
relaxation other than radiative decay or
due to inhomogeneous (e.g., Doppler) broadening. We also require that
the $|2\rangle-|4\rangle$ transition, used for high-contrast EIT, has
a very narrow linewidth.
We solve for the steady-state values of
the density matrix elements $\rho_{34}$ and $\rho_{21}$ in a linear 
approximation in which the probe field components $\Eb$ and $\Bb$ are 
assumed to be small variables:

\be
\rho_{34} =  \alpha_{EE}\Eb+\alpha_{EH}\Bb, \qquad \rho_{21} =
\alpha_{HE}\Eb+\alpha_{HH}\Bb .
\label{response}
\ee
Here $\alpha_{EE}$ and $\alpha_{HH}$ are the polarizabilities and 
$\alpha_{EH}$ and $\alpha_{HE}$ are the chirality parameters. From 
Eq.~(\ref{response}) we determine the ensemble polarization $\Pb = N 
d_{34} \rho_{34}$
and magnetization $\Mb = N \mu_{21} \rho_{21}$, where
  $N$ is the density of radiators (atoms, molecules,
excitons, etc.) and $d_{34}$ and $\mu_{21}$ are
the probe field transition electric
and magnetic dipole moments.
%%%%%%%%%%%%%%%%
% Explicit expressions for the
% polarizabilities ($\alpha_{EE}$, $\alpha_{HH}$) and chirality
% parameters ($\alpha_{EH}$, $\alpha_{HE}$)
% are given in the methods section.
%%%%%%%%%%%%%%%%
Fig.~\ref{fig2a} shows typical
calculated spectra for the electric and magnetic
polarizabilities as well as the
chirality parameters. The induced reduction of the electric
polarizability $\alpha_{EE}$ and thus of the absorption
on resonance is apparent. Likewise one recognizes a resonantly enhanced
imaginary chirality $\alpha_{HE}$ while the other chirality
$\alpha_{EH}$ and the magnetic polarizability $\alpha_{HH}$ remain
very small.

%%%%%%%%%%%%%%%%%%%%%%%%%%%%%%%%%%%%%%%%%%%%%%%%%%%%%%%%%%%%%%%%%%%%%%%%%%
%%%%%%%%%%%%%%%%%%%%%%%%%%%%%%%%%%%%%%%%%%%%%%%%%%%%%%%%%%%%%%%%%%%%%%%%%%

Since we are interested in values for the refractive index
substantially different from unity, we must take into
account local field corrections.
Due to the induced chirality, the probe magnetic field couples to the
media by a
factor $\simeq$ 137 times stronger
than in the case of non-chiral media.
Therefore we include both electric and
magnetic local field effects \cite{Cook,Kaestel07} to first order by replacing
the macroscopic field values
$\Eb$ and $\Bb$ in Eq.~(\ref{response})
with the local (i.e., microscopic) fields $\Eb_\text{m} = \Eb + 
\Pb/3\varepsilon_0$
and $\Bb_\text{m}/\mu_0 = \Hb_\text{m} = \Hb +\Mb/3$.
%%%%%%%%%%%%%%%%
% (Higher-order
% corrections, e.g., modification of radiative decay rates,
% are not necessary for high-contrast EIT.)
%%%%%%%%%%%%%%%%
We then use the polarizabilities and chirality parameters to
calculate the permittivity $\varepsilon$, permeability $\mu$, and
chirality coefficients $\xi_{EH}$ and $\xi_{HE}$ with local
field corrections, and determine the refractive index from
Eq.~(\ref{refind}).
%%%%%%%%%%%%%%%%
% (For details see the methods section.)
%%%%%%%%%%%%%%%%

\begin{figure}[tb]
      \begin{center}
        \includegraphics[width=5.6cm]{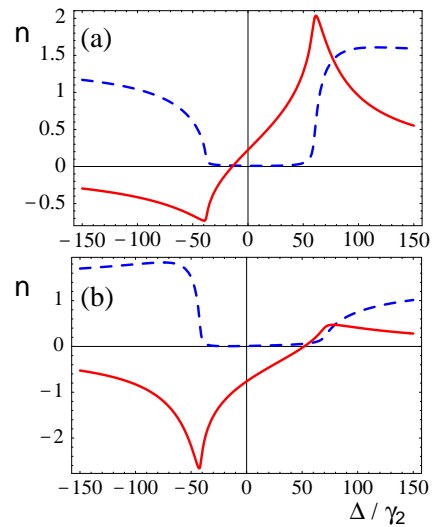}
        \caption{(color online) Real ({\color{red} solid}) and 
imaginary ({\color{blue} dashed})
parts of the refractive index
% , including local field effects,
% as a function of the probe field detuning $\Delta=-\Delta_E=-\Delta_B$
% relative to the radiative decay rate $\gamma_2$ from level
% $|2\rangle$,
for a
density of (a) $N=5\cdot 10^{16}\text{cm}^{-3}$ and (b) $N=5\cdot
10^{17}\text{cm}^{-3}$.
% Other system parameters are given in the text
% and all coupling fields are resonant.
}
        \label{fig2}
      \end{center}
\end{figure}

As an example, Fig.~\ref{fig2}(a) shows the calculated real and imaginary
refractive index as a function of probe field detuning for a density
of $N=5\cdot 10^{16}\text{cm}^{-3}$ and using the realistic parameter
values $\gamma_2=10^3\text{s}^{-1}$, $\Omega_c=10^4e^{i\pi/2}\gamma_2$,
$\Omega_1=\Omega_2=10^6\text{s}^{-1}$,
$\gamma_3=\gamma_5=(137)^2\gamma_2$, $\gamma_4\simeq0$, and
$\gamma_P=10^4\gamma_2$.
The transition dipole matrix elements are related to the radiative decay rates
via
$d_{34}(\mu_{21}/c)=\sqrt{3\pi\varepsilon_0\gamma_3(\gamma_2)\hbar 
c^3/\omega^3}$
for
a typical optical wavelength of $600$nm.
%%%%%%%%%%%%%%%%
% (See the methods
% section for the definition of the above parameters; and note that the
% detunings and Rabi frequencies are given as angular
% frequencies.)
%%%%%%%%%%%%%%%%
We find
substantial negative refraction and minimal absorption for this density,
which is about a factor of $10^2$ smaller than the density needed without
taking chirality into account \cite{Oktel04}.
It should also be noted that the magnetic permeability differs by less than
10\% from unity for the density considered here. Therefore, the
negative refraction shown in
Fig.~\ref{fig2}(a) is clearly a consequence of chirality induced by
quantum interference.
%%%%%%%%%%%%%%%%%%%%%%%%%%%%%%%%%%%%%%%%%%%%%%%%%%%%%%%%%%%%%%%%%%%%%%%%%%
%%%%%%%%%%%%%%%%%%%%%%%%%%%%%%%%%%%%%%%%%%%%%%%%%%%%%%%%%%%%%%%%%%%%%%%%%%

\begin{figure}[tb]
      \begin{center}
        \includegraphics[width=7cm]{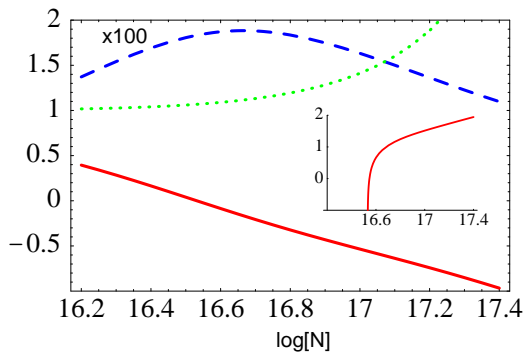}
        \caption{(color online) Real ({\color{red} solid}) and 
imaginary ({\color{blue}
dashed}, $\times 100$) parts of the
refractive index, as well as the real part of the permeability
({\color{green} dotted}), as a
function of the logarithm of the density $N$, at a frequency slightly
below resonance ($\Delta=-25\gamma_2$).
% and including local field effects.
Inset: $\log\left|{\Re(n)}/{\Im(n)}\right|$ as a
function of $\log[N]$.
% Other system parameters are the same as for
% Fig.~\ref{fig2} and all coupling fields are resonant.
}
        \label{fig4}
      \end{center}
\end{figure}

For larger $N$ the optical response of the medium increases. For
example, Fig.~\ref{fig2}(b) shows the refractive index for $N=5\cdot
10^{17}\text{cm}^{-3}$. % with otherwise unchanged parameters.
As expected the refraction (real part of $n$) reaches increasingly
large negative values.
Remarkably, we find that the absorption
reaches a maximum and then decreases with increasing density.
This effect is illustrated in  Fig.~\ref{fig4} where we show the
real and imaginary parts of the refractive index as well as the real part
of $\mu$ (including local-field effects) as functions of the density. This
peculiar behaviour is due to the term
proportional to $(\xi_{EH}+\xi_{HE})^2$ in Eq. (\ref{refind}),
which is nonzero in a broadened medium where $\xi_{EH}\ne-\xi_{HE}$
and can have
a positive imaginary part and thus can partially compensate the
imaginary part of $\varepsilon\mu$ and due to local field corrections 
\cite{Kaestel07}. As a consequence the
refraction/absorption ratio, shown in the inset of Fig.~\ref{fig4},
continues to increase with density and reaches rather large values
on the order of 10$^2$. These results should be contrasted to previous
theoretical proposals and experiments on negative refraction in the optical
regime, for which $-\Re[n]/\Im[n]$ %the refraction/absorption ratio
is typically on the
order of unity.

%%%%%%%%%%%%%%%%%%%%%%%%%%%%%%%%%%%%%%%%%%%%%%%%%%%%%%%%%%%%%%%%%%%%%%%%%%
%%%%%%%%%%%%%%%%%%%%%%%%%%%%%%%%%%%%%%%%%%%%%%%%%%%%%%%%%%%%%%%%%%%%%%%%%%

\begin{figure}[tb]
      \begin{center}
        \includegraphics[width=6cm]{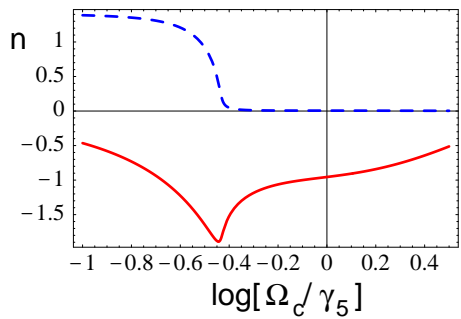}
        \caption{(color online) Real ({\color{red} solid}) and 
imaginary ({\color{blue}
dashed}) parts of the refractive index
as a function of the coupling field Rabi-frequency
$\Omega_c$
% relative to the radiative decay rate $\gamma_3$,
for $N=3.5\cdot10^{17}$ cm$^{-3}$ and $\Delta=-25\gamma_2$.
% $\Delta=-\Delta_B=-\Delta_E=-25\gamma_2$.
% Other system parameters are the same as for
% Figs.~\ref{fig2} and \ref{fig4} and all coupling fields are resonant.
}
        \label{fig-tuning}
      \end{center}
\end{figure}

It has been pointed out by Smith {\it et al.} and Merlin
\cite{Schmith03} that the
realization of sub-diffraction-limit imaging with a lens of thickness
$d$ and resolution $\Delta x$
requires an extreme fine tuning of the index of refraction to the value
$n=-1$ with accuracy $\Delta n = 1-\exp\left\{-\frac{\Delta x}{2\pi 
d}\right\}$.
The quantum interference scheme presented here provides a handle for
such fine tuning. For example, as shown in Fig.~\ref{fig-tuning}, the
real part of the refractive index can be fine tuned by relatively
coarse adjustments of the strength of the coupling field $\Omega_c$.

%%%%%%%%%%%%%%%%%%%%%%%%%%%%%%%%%%%%%%%%%%%%%%%%%%%%%%%%%%%%%%%%%%%%%%%%%%
%%%%%%%%%%%%%%%%%%%%%%%%%%%%%%%%%%%%%%%%%%%%%%%%%%%%%%%%%%%%%%%%%%%%%%%%%%

As with EIT and its applications,
electromagnetically-induced chiral negative refraction should be
realizable in a wide range of atomic, molecular, and condensed matter
systems. Already, a basic form of electromagnetically-induced
chirality has been realized in Rb vapor \cite{Scully-chirality}. In
preparation for further experimental investigations we are currently
performing a detailed assessment of systems such as metastable neon,
rare-earths (dysprosium and others), as well as donor-bound electrons
and bound excitons in
semiconductors, all of which can have level structures and interactions
analogous to those of the modified level scheme of Fig.~\ref{fig1}(b).
%%%%%%%%%%%%%%%%
% Results from these systems will be reported in future
% publications.
%%%%%%%%%%%%%%%%

In conclusion, we showed that
quantum interference effects can lead to a large induced chirality
under realistic conditions, and thus enable tunable
negative refraction with minimal absorption. Such electromagnetically-induced
chiral negative refraction should be applicable to a wide array of
systems in the optical regime, including atoms, molecules, quantum
dots, and excitons.

\begin{acknowledgments}
M.F. and J.K. thank the Institute for Atomic, Molecular and
Optical Physics at the Harvard-Smithsonian Center for Astrophysics and
the Harvard Physics Department for their hospitality and support.
R.W. thanks D. Phillips for useful discussions.
J.K. acknowledges financial support by the Deutsche Forschungsgemeinschaft
through the GRK 792 ``Nichtlineare Optik und Ultrakurzzeitphysik''.
S.Y. thanks the Research Corporation for support.
\end{acknowledgments}

\def\etal{\textit{et al.}}


\begin{thebibliography}{99}

\bibitem{Veselago68}
V. G. Veselago, Sov. Phys. Usp. {\bf 10}, 509 (1968).

\bibitem{Pendry00}
J. B. Pendry, Phys. Rev. Lett. {\bf 85}, 3966 (2000).

\bibitem{Pendry99}
J. B. Pendry \etal, IEEE Trans. Micro. Theory Tech. {\bf 47}, 2075 (1999).

\bibitem{Smith00}
D. R. Smith \etal, Phys. Rev. Lett. {\bf 84}, 4184 (2000); R. Shelby, 
D. R. Smith, and S. Schultz, Science {\bf 292}, 77 (2001).

\bibitem{Yen04}
T. J. Yen \etal, Science {\bf 303}, 1494 (2004).

\bibitem{Linden04}
S. Linden \etal, Science {\bf 306}, 1351 (2004); C. Enkrich \etal, 
Phys. Rev. Lett. {\bf 95}, 203901 (2005).

\bibitem{Parimi04}
P. V. Parimi \etal, Phys. Rev. Lett. {\bf 92}, 127401 (2004).

\bibitem{Berrier04}
A. Berrier \etal, Phys. Rev. Lett. {\bf 93}, 073902 (2004).

\bibitem{Lu05}
Z. Lu \etal, Phys. Rev. Lett. {\bf 95}, 153901 (2005).

\bibitem{Fleischhauer05}
M. Fleischhauer, A. Imamoglu, and J. P. Marangos, Rev. Mod. Phys. 
{\bf 77}, 633 (2005).

\bibitem{Scully-chirality}
V. A. Sautenkov \etal, Phys. Rev. Lett. {\bf 94}, 233601 (2005).

\bibitem{Schmith03}
D. R. Smith \etal, Appl. Phys. Lett. {\bf 82} 1506 (2003);
R. Merlin, Appl. Phys. Lett. {\bf 84}, 1290 (2004).

\bibitem{Pendry04}
J. B. Pendry, Science {\bf 306}, 1353 (2004).

\bibitem{Harris-NLO-EIT}
S. E. Harris, J. E. Field, and A. Imamoglu, Phys. Rev. Lett. {\bf 
64}, 1107 (1990).

\bibitem{Stoicheff-Hakuta}
K. Hakuta, L. Marmet, and B. P. Stoicheff, Phys. Rev. Lett. {\bf 66}, 
596 (1991).

\bibitem{Oktel04}
M. \"O. Oktel, and \"O. E. M\"ustecaplioglu, Phys. Rev. A {\bf 70}, 
053806 (2004).

\bibitem{Thommen-PRL-2006}
Q. Thommen, and P. Mandel, Phys. Rev. Lett. {\bf 96}, 053601 (2006).

\bibitem{footnote} In \cite{Thommen-PRL-2006} a negative
value for Im$[n]$ is given, corresponding to an amplifying medium;
however this result is due to
a calculational error, see J. K\"astel, \& M. Fleischhauer, Phys. 
Rev. Lett. {\bf 98}, 069301 (2007).

\bibitem{Cook}
D. M. Cook, {\it The Theory of the Electromagnetic Field},
Prentice-Hall (New Jersey).

\bibitem{Kaestel07}
For a rigorous derivation of generalized Clausius-Mossotti relations 
in magneto-dielectric media see:
J. K\"astel, G. Juzeliunas, and M. Fleischhauer (in preparation)

\end{thebibliography}
\end{document}